\documentclass{birkjour}

\usepackage[pdftex,
            pdfauthor={Ross N. Greenwood},
            pdftitle={On Computable Transformations of Geometric Expressions in Quantum Theory}]{hyperref}
            
\usepackage{ragged2e}

\usepackage{graphicx}        
\usepackage{multicol}        
\usepackage[bottom]{footmisc}

\makeindex

\usepackage{amsmath,amssymb,amsfonts,mathabx,enumitem,graphicx,subfigure,mathrsfs}
\usepackage{setspace}

\newcommand{\dx}{\mathrm d}

\newcommand{\bra}[1]{\left\langle #1 \right\rvert}
\newcommand{\ket}[1]{\left\lvert #1 \right\rangle}

\newcommand{\fslash}[1]{\setbox 0 \hbox{$#1$} \rlap{\hbox to \wd 0 {\hss/\hss}} \box 0}
\newcommand{\matzero}{\setbox 0 \hbox{$\phantom{-1}$} \rlap{\hbox to \wd 0 {\hss0\hss}} \box 0}

\DeclareMathOperator{\Tr}{Tr}

\newcommand{\hatcal}[1]{\hat{\mathcal #1}}
\newcommand{\otimesassign}{\setbox 0 \hbox{${}\otimes{}$} \rlap{\hbox to \wd 0 {\hss$\longleftarrow$\hss}} \box 0}

\DeclareMathSymbol{\varPi}{\mathalpha}{operators}{"05}
\DeclareMathSymbol{\varLambda}{\mathalpha}{operators}{"03}

\renewcommand{\theequation}{\arabic{equation}}
\newcommand{\ClR}{\mathcal C\ell_3(\mathbb R)}

\begin{document}

    \title[On Computable Geometric Expressions in Quantum Theory]{On Computable Geometric Expressions \\in Quantum Theory}
    
    \author{Ross N. Greenwood}
    \address{
    Department of Physics \\
    University of California Santa Cruz \\
    1156 High Street \\
    Santa Cruz, CA 95064
    }
    \email{ross.greenwood@ucsc.edu}
	
    \begin{abstract}
    Geometric Algebra and Calculus are mathematical languages encoding fundamental geometric relations that theories of physics seem to respect. 
    We propose {criteria given} which statistics of expressions in geometric algebra are computable in quantum theory, in such a way that preserves their algebraic properties.
    They are that one must be able to arbitrarily transform the basis of the Clifford algebra, via multiplication by elements of the algebra that act trivially on the state space; 
    all such elements must be neighbored by operators corresponding to factors in the original expression and not the state vectors.
    We explore the consequences of these criteria for a physics of dynamical multivector fields.
    \end{abstract}
    
    
    \maketitle
    
    \section{Introduction}

Multivectors are elements of a Clifford algebra atop $\mathbb R^n$ that may represent formal sums of oriented geometric extents in a generalized tangent space
(see Appendix A for a brief overview, and e.g.\ \cite{hest1,doran_lasenby_2003,macd1} for in-depth coverage).
In \cite{Hestenes:2008qj,Lasenby2019,2005AnPhy.317..383F}, a distinction is made between multivectors whose components transform under rotations via one- versus two-sided Clifford multiplication:
\begin{equation} \psi \to S \psi \quad\text{vs.}\quad \mathcal F \to S \mathcal F S^\dag  \end{equation}
where $S \in \left\{ \exp(\tfrac14 \epsilon_{ijk} \theta_i \mathbf e^j \mathbf e^k) \mid \theta_i \in \mathbb R \right\}$ is a {\it rotor} and $^\dag$ denotes a reverse involution. 
Though the two-sided transformation of the latter is standard in geometric algebra, it is suggested in \cite{Hestenes:2008qj} that the former are the more fundamental objects because the latter can be constructed from the former
\begin{equation} \psi \psi^\dag \to S (\psi \psi^\dag) S^\dag \end{equation}
but not vice versa.

In order for expectation values and higher moments of random variable expressions in geometric algebra to be ``computable'' in terms of state vectors in the probabilistic framework of quantum theory, one must remain free to distort the basis of the Clifford algebra 
by applying linear operators---the rotors introduced above.  
Rotors can only act on elements of the geometric algebra, which must first be appended to the state vector by the action of a field operator (or a non-field  ``catalog'' operator).
Since rotors cannot be allowed to act directly on state vectors living in a separate Hilbert space, it would seem that multivectors transforming by one-sided multiplication (like $\psi$ above) must occupy the outer positions\footnote{If the expression in geometric algebra is $ABC$, and its expectation value given $\ket\Psi$ is computed as $\mathrm E[ABC] = \bra{\Psi}\hat A \hat B \hat C \ket\Psi$, then $\hat A$ and $\hat C$ occupy ``outer positions.''} adjacent to the state vectors in any expression
that is computable by the above criterion.

We 
present a scheme in which dynamical multivector {fields} whose components {\it for all particular intents} transform by left-multiplication alone may be constructed from those transforming natively with the standard two-sided rule.  These fields are coupled in one of two configurations to a gauge-like connection that transforms non-covariantly under basis rotations  applied to fibers of the Clifford bundle.
We argue that expressions of multivector-valued fields and their derivatives must conform to these configurations to be computable in the linear associative framework of quantum theory.
We draw a comparison between this scheme and the fermion-gauge field couplings prescribed by the Weinberg-Salam electroweak model.
    \section{Mixing Geometric Algebra and State Vectors}
\label{sec1}

Assembling a generic {multivector} $\mathcal F$ in the geometric algebra of three spatial dimensions, $\mathcal C\ell_3(\mathbb R)$, we have
\begin{align} \mathcal F &\equiv \, b_0 \mathbf 1 + b_i \mathbf e^i + b_{jk} \mathbf e^j \mathbf e^k + b_{123} \mathbf e^1 \mathbf e^2 \mathbf e^3 \\ 
&\mapsto b_0 \varsigma^0 + b^i \varsigma^i + i \epsilon_{ijk} b_{jk} \varsigma^i + i b_{123} \varsigma^0 \label{Fdef}  \end{align}
Here we take $\mathbf e^i$ to be orthonormal vector basis elements of that algebra, with matrix representations
$\varsigma^i$
satisfying $\varsigma^1 \varsigma^2 = i \varsigma^3$ and $-i \varsigma^1 \varsigma^2 \varsigma^3 = I_d \equiv \varsigma^0$;
all coefficients are real-valued.
Consider expressions taking the form
\begin{equation} \left\langle \mathcal F^\dag \mathcal O \mathcal F \right\rangle_0 = \left\langle \mathcal F \mathcal F^\dag \mathcal O \right\rangle_0 = \left\langle  \mathcal O \mathcal F \mathcal F^\dag \right\rangle_0 \label{fof_ga} \end{equation}
for some multivector element $\mathcal O$, where $\langle\cdot\rangle_0$ denotes the grade-0 projection. Adopting a matrix representation, the grade-0 projection is obtained up to a constant factor as the trace of the Hermitian part.
Using $S^\dag = S^{-1}$ and the cyclic property of $\langle\cdot\rangle_0$, the outer transformation operators acting on $\mathcal F$ cancel out, leaving the one-sided effective transformation $\mathcal F \to S \mathcal F$.
\begin{gather} 
\left\langle \mathcal F^\dag \mathcal O \mathcal F \right\rangle_0 \to \left\langle (S \mathcal F S^\dag)^\dag \mathcal O' (S \mathcal F S^\dag) \right\rangle_0 = \left\langle \mathcal F^\dag S^\dag \mathcal O' S \mathcal F \right\rangle_0 \label{cyclic1}
\end{gather}

Now suppose one exports information about the configuration of $\mathcal F$ to a state vector $\ket\Psi$. In place of ${\mathcal F}$ one substitutes a linear
operator $\hatcal F$ mapping between the Hilbert space of states $\mathscr H$ and a composite space $(\mathcal C\ell_3(\mathbb R), \mathscr H)$ containing the geometric algebra, e.g.\
\begin{align} 
\hat{\mathcal F} &= b_0 \varsigma^0 \overset\leftarrow\otimes \hat a_0(b_0) + b_i \varsigma^i \overset\leftarrow\otimes \hat a_i(b_i) + b_{jk} \varsigma^j \varsigma^k \overset\leftarrow\otimes \hat a_{jk}(b_{jk}) + \cdots \\
\intertext{or $\check{\mathcal F} \, \cdot : (\mathcal C\ell_3(\mathbb R), \mathscr H) \mapsto (\mathcal C\ell_3(\mathbb R), \mathscr H)$ mapping among elements of that composite space}
\check{\mathcal F} &= b_0 \varsigma^0 \otimes \hat a_0(b_0) + b_i \varsigma^i \otimes \hat a_i(b_i) + b_{jk} \varsigma^j \varsigma^k \otimes \hat a_{jk}(b_{jk}) + \cdots \end{align}
Here $\hat a_{\{\cdot\}} : \mathscr H \to \mathscr H$ transform vectors in the Hilbert space, and $\overset\leftarrow\otimes$ denotes that $\hatcal F $ maps between $\mathscr H$ on the right and a composition $(\mathcal C\ell_3(\mathbb R), \mathscr H)$ on the left.
We refer to $\hat{\mathcal F}$ as a \textit{catalog} operator, by analogy to a store catalog that maps codes in an inventory record to representations of concrete objects; elements of $(\mathcal C\ell_3(\mathbb R), \mathscr H)$ are multivectors appended to state vectors.
(From this point forward, a hat (\;$\hat{}$\;) denotes a rectangular operator that maps between the Hilbert space and a larger composite space; a caron ($\;\check{}\;$) denotes a square operator that acts non-trivially on $\mathscr H$, but maps between identical spaces; and operators like $S$ that act trivially on $\mathscr H$ have no ornament.)

One can transform each domain of the catalog operator separately
\begin{equation*} \check{\mathcal F} \to \begin{cases}
b_0 S \varsigma^0 S^\dag \overset\leftarrow\otimes \check a_0(b_0) + b_i S \varsigma^i S^\dag \overset\leftarrow\otimes \check a_i(b_i) + \cdots & \text{(Clifford basis)} \\
b_0 \varsigma^0 \overset\leftarrow\otimes U \check a_0(b_0) U^\dag + b_i \varsigma^i \overset\leftarrow\otimes U \check a_i(b_i) U^\dag + \cdots & \text{(Hilbert space map)} \end{cases} \end{equation*}
with the state vector 
transforming under a unitary representation of $SO(3)$ as $\ket\Psi \to U\! \ket\Psi$.
This work relies on an insistence that one be able
to compute expectation values of transformed expressions after a change of the Clifford basis in terms of the original catalog operators, in a manner consistent with the linear formulation of quantum theory.  We denote the transformation acting on the Clifford basis as simply
\begin{equation} \check{\mathcal F} \to S \check{\mathcal F} S^\dag 
\end{equation}
Note that the rectangular hatted form $\hatcal F \,\cdot : \mathscr H \mapsto (\mathcal C\ell_3(\mathbb R),\mathscr H)$ cannot be acted upon from the right by any operator not merely mapping $\mathscr H \mapsto \mathscr H$, without changing the input space of the resulting operator.

Computing the expectation value of the quantity \eqref{fof_ga} for a given state $\ket\Psi$ after a transformation of the Clifford {basis} (not of $\mathscr H$), we are free to omit transformation operators that cancel out in \eqref{cyclic1}
\begin{equation} \bra\Psi \Tr_\varsigma \left[ {\check{\mathcal F}^\dag \check{\mathcal O} \hat{\mathcal F}} \ket\Psi \right] + \text{h.c.} \to \bra\Psi \Tr_\varsigma \left[ {\check{\mathcal F}^\dag S^\dag \check{\mathcal O}' S \hat{\mathcal F}} \ket\Psi \right] + \text{h.c.}
\end{equation}
Here $\Tr_\varsigma : (\mathcal C\ell_3(\mathbb R), \mathscr H) \mapsto \mathscr H$ denotes a partial trace over the matrix representation of the geometric algebra, and $\mathcal O \to \mathcal O'$ under basis rotation.
We henceforth suppress addition of the Hermitian conjugate when computing grade-0 projections in the matrix representation.

From a geometric standpoint, nothing changes if one cycles the order of multiplication as in \eqref{fof_ga} (while accordingly reassigning the ladder operators 
$\check a_{\{\cdot\}} : \mathscr H \mapsto \mathscr H$),
after which the transformation becomes
\begin{equation} \bra\Psi \Tr_\varsigma \left[ \check{\mathcal O} \check{\mathcal F} \hat{\mathcal F}^\dag \ket\Psi \right] \to \bra\Psi \Tr_\varsigma \left[ \check{\mathcal O'} (S \check{\mathcal F} S^\dag) (S \hat{\mathcal F}^\dag S^\dag) \ket\Psi \right] \label{blahh} \end{equation}
If one takes associativity of the operations in brackets literally, one must allow for the interpretation of the transformation operator $S^\dag$ as acting directly on the state vector to the right.
But $\ket{\Psi}$
does not contain an element of the geometric algebra 
in the representation on which $S$ is to act.
Nor does $S$ constitute a map between the Hilbert space and a larger composite space containing the geometric algebra in the representation $\varsigma$, as does $\hat{\mathcal F}$.

If we are to be free in principle to evaluate $\eqref{fof_ga}$ with any ordering, then a basis transformation cannot {require} multiplication by $S^\dag$ from the right on the right-most catalog operator $\hat{\mathcal F}^\dag$ in \eqref{blahh}, between the latter and the state $\ket{\Psi}$.
Only if $\mathcal O$ simultaneously transforms as $\mathcal O \to \mathcal O' = S \mathcal O S^\dag$, or commutes with all $S$, is the transformed expression computable.  
\begin{equation} \bra\Psi \Tr_\varsigma \left[ S \check{\mathcal O} S^\dag S \check{\mathcal F} \hat{\mathcal F}^\dag S^\dag \ket\Psi \right] \equiv \bra\Psi \Tr_\varsigma \left[ \check{\mathcal O} \check{\mathcal F} \hat{\mathcal F}^\dag \ket\Psi \right] \end{equation}
This amounts to a constraint on the geometric quantities about which information can be retrieved from state vectors.
By this reasoning, for example, the quantity
$ \langle \mathcal F^\dag \mathbf e^i \mathcal F \rangle_0 $ ---where $\mathbf e^i$ is a constant reference vector that is left invariant under basis rotations---would not be admissible.
(For multivector \textit{fields}, we will find that including a right-acting derivative operator in the mix imposes an order of multiplication that cannot be permuted without changing the content of the expression, allowing constant reference elements $\mathbf e^k$ to appear alongside $\partial_k$.)

Having verified that the transformed expression in terms of geometric algebra is free of outer basis transformation operators, we are free to discard the curious notation adopted above.
We can substitute a vector-only notation---replacing $\hat{\mathcal F}$ with a set of catalog operators mapping $\ket\Psi$ to a tensor product space containing columns of the matrix representation---with no loss of fidelity.
\begin{equation} \mathrm E \left[ \left\langle \mathcal F^\dag \mathcal O \mathcal F \right\rangle_0 \right] \sim {\textstyle \sum_A} \, \bra\Psi \hat{\psi}_A^\dag \check{\mathcal O} \hat{\psi}_A \ket\Psi, \quad A \in \{1,\dots,\mathrm{dim}\; \varsigma^0 \} \label{vectorized} \end{equation}
If the $2\times2$ complex representation is adopted, then the $\psi_A$ for $A \in \{1,2\}$ are the usual 2-component Weyl spinors, \emph{effectively} transforming as $\psi_A \to S \psi_A$.
    
\section{Dynamical Multivector Fields}

Let us attempt to construct a physics of elementary multivector fields, transforming with the standard two-sided Clifford multiplication rule.  A candidate free Hamiltonian for such a theory is given by
\begin{equation} \dx H = \left\langle \dx^3 \mathbf x \, \mathcal H(\mathbf x) \right\rangle_0 =  \left\langle \dx^3 \mathbf x \, \mathcal F(\mathbf x)^\dag \mathbf e^k \partial_k \mathcal F(\mathbf x) \right\rangle_0 \label{foga} \end{equation}
where $\langle\cdot\rangle_0$ again denotes the grade-0 projection, and the 3-volume element $\dx^3 \mathbf x \equiv \dx^3 \mathbf x(\mathbf x)$ is taken to be a homogeneous multivector field of grade 3.  It represents the oriented volume of an infinitesimal voxel at $\mathbf x$, against which the orientation of the multivector integrand is evaluated.
\begin{equation} \textstyle \dx^3 \mathbf x \equiv \bigwedge_{k=1}^3 \dx x^k(\mathbf x) \, \mathbf e_k = S \, \dx^3 \mathbf x \, S^\dag \end{equation}
The $\mathbf e^k$ in \eqref{foga} are constant reference vector elements of the geometric algebra; they do not transform with the basis of the local Clifford algebra as does the field.
The Hamiltonian is the integral of \eqref{foga} over all of space;
for now we only consider configurations for which the integral is positive.

\subsection{Rotations}

For multivector fields, two transformations are involved in a change of reference frames: one of the coordinates on the manifold and another of the basis of multivectors spanning the Clifford algebra at each event
(the union of such spaces forms the Clifford bundle).
For brevity, we will often refer to the latter as simply {\it basis transformations}.

It is convenient to adopt a {\it holonomic} basis for the Clifford bundle, in which the vector basis elements of the algebra at each point in the manifold are taken to align with the directions along which the spatial coordinates increase.
However, non-holonomic bases are admissible and should describe the same physics when reflected in the state vector.
The Hamiltonian must be invariant under complementary coordinate and basis transformations, but not under transformations between holonomic and non-holonomic bases. (The latter change the geometric interpretation of the information contained in the state vector (e.g. spin orientation relative to field gradient), and so must correspondingly give a different time evolution for that new physical system.)

Let new coordinates $\bar {\mathbf x}$ relate to the old as $\bar {\mathbf x} = \varLambda \mathbf x$, and let the multivector $S = \exp ({\tfrac14 \epsilon_{ijk} \theta_i \mathbf e^j \mathbf e^k})$ 
operate on elements of the geometric algebra as $S({}\cdot{})S^\dag$ to enact a proper basis rotation.
Under the combined rotation
\begin{equation} \dx H(\mathbf x) \to \left\langle (S \, \dx^3 \mathbf x \, S^\dag) (S \mathcal F(\varLambda^{-1} \mathbf x)^\dag S^\dag) \,\mathbf e^k \partial_{k} (S \mathcal F(\varLambda^{-1} \mathbf x) S^\dag) \right\rangle_0 \label{foga_trans} \end{equation}
where we assume $\dx^3 \mathbf x(\varLambda^{-1}\mathbf x) \equiv \dx^3 \mathbf x(\mathbf x)$. When the rotation is homogeneous ($S(\mathbf x) = S$), with the cyclic property of $\langle \cdot \rangle_0$ this becomes
\begin{equation} \dx H(\mathbf x) \to \dx H_{\text{hom}}(\mathbf x) = \left\langle \dx^3 \mathbf x \, \mathcal F(\varLambda^{-1} \mathbf x)^\dag S^\dag \mathbf e^k S \partial_{k} \mathcal F(\varLambda^{-1} \mathbf x) \right\rangle_0 \label{foga_trans2} \end{equation}
If the basis transformation coded by $S$ is not homogeneous (e.g. transforming between holonomic bases corresponding to Cartesian and curvilinear coordinates), then we get additional terms from the transformation
\begin{multline} \dx H(\mathbf x) \to \dx H_{\text{hom}}(\mathbf x) + \left\langle \dx^3 \mathbf x \, \mathcal F(\varLambda^{-1} \mathbf x)^\dag S^\dag \mathbf e^k (\partial_k S) \mathcal F(\varLambda^{-1} \mathbf x) \right\rangle_0 + \\ \left\langle \dx^3 \mathbf x \, \mathcal F(\varLambda^{-1} \mathbf x)^\dag S^\dag \mathbf e^k S \mathcal F(\varLambda^{-1} \mathbf x) (\partial_k S^\dag) S \right\rangle_0 \label{foga_trans3} \end{multline}
This transformed Hamiltonian is of course still well defined as a sequence of operations involving multivectors followed by a grade projection.  

But what happens when we try to compute statistics of this Hamiltonian in terms of data externalized to a state vector, while taking seriously the requirement that the probability theory be linear and associative?
We promote $\mathcal F(\mathbf x)$ to a field operator $\hat{\mathcal F}(\mathbf x)$, which acts on the state vector $\ket{\Psi}$ serving as a dictionary between the Hilbert space and sections of the Clifford bundle.
We can cast the derivatives as right-acting linear operators,
requiring only that $\mathcal F$ appears alone to the right of the derivative operator in the untransformed expression (and so fixing the order of multiplication).

When the transformation of the field values is homogeneous, we can use the cyclic property of the grade-0 projection in \eqref{foga}, to cancel the left-most $S$ and the right-most $S^{\dag}$
in the basis transformed version \eqref{foga_trans}.
We are free to omit those operators that do not survive in the transformed expression
\begin{equation} \bra\Psi \!\hat{H}\! \ket\Psi
\overset{\partial S = 0}{- \!\!\! \longrightarrow} 
\bra\Psi \Tr_\varsigma \int \dx^3 \mathbf x \, \hat{\mathcal F}^\dag S^\dag \varsigma^k S \overset\rightarrow{\partial_k} \hat{\mathcal F} \ket{\Psi} \end{equation}
leaving a field operator at the right-most position mapping the state to an element of a composite space $(\mathcal C\ell_3(\mathbb R),\mathscr H)$ that includes a copy of the geometric algebra in the representation specified by the reference vector $\varsigma^k$. 
If the basis transformation is not uniform, then we get a term
\begin{equation} \bra\Psi \!\hat{H}\! \ket\Psi \to \cdots {}+{} \bra\Psi \Tr_\varsigma \int S \, \dx^3 \mathbf x \, \hat{\mathcal F}^\dag S^\dag \varsigma^k S \hat{\mathcal F} \overset\rightarrow{\partial_k} S^{\dag} \ket\Psi \end{equation}
in which we would have to allow for the interpretation of $S^{\dag}$ as acting on $\ket{\Psi}$.
But as before, the basis transformation operator $S^\dag$ cannot be allowed to act directly on the bare quantum state (or on $\hatcal F$ from the right), since an element of the Clifford algebra in the representation consistent with $\varsigma^k$ has not yet been appended to the state vector.
The Hamiltonian must be constructed such that all \emph{outer} basis transformation operators (OBTOs) can be omitted without consequence.

In order to accommodate a kinetic term resembling \eqref{foga} in a computable quantum field theory, one would have to introduce an auxiliary grade-2 field with a coordinate index, whose components transforms non-covariantly under basis rotations:
\begin{equation} 
\mathcal W_k(\mathbf x) \to \begin{cases}
S(\mathbf x) \mathcal W_k(\mathbf x) S(\mathbf x)^{\dag} - S(\mathbf x) \partial_k S(\mathbf x)^{\dag} & \text{(Clifford basis)} \\
\varLambda^k_{\phantom k {k'}} \mathcal W_{k}(\varLambda^{-1} \mathbf x) & \text{(coordinates)} 
\end{cases} \label{Wdef} \end{equation}
Using \eqref{Wdef} 
there are {two} available modifications to \eqref{foga} that resolve the issue:
\begin{equation} \mathcal H_\pm(\mathbf x) = \mathcal F^\dag \mathbf e^k \partial_k \mathcal F + 
 \mathcal F^\dag \mathbf e^k [\mathcal F, \mathcal W_k] \pm \mathcal F^\dag \mathbf e^k \mathcal W_k \mathcal F \label{blah} \end{equation}
Using $S \partial_\mu S^\dag = - S^\dag \partial_\mu S$,
the second term first removes all dependence on $\partial_k S$, and then the third term restores an inner dependence up to a sign:
\begin{align} 
\mathcal H_+(\mathbf x) &\to 
\mathcal F^\dag \mathbf e^{k'} S^\dag \partial_{k} (S \mathcal F) + \mathcal F^\dag \mathbf e^{k'} \mathcal F \mathcal W_k \\
 \mathcal H_-(\mathbf x) &\to \mathcal F^\dag \mathbf e^{k'} S \partial_k (S^\dag \mathcal F) + \mathcal F^\dag \mathbf e^{k'} \mathcal F \mathcal W_k - 2 \mathcal F^\dag \mathbf e^{k'} \mathcal W_k \mathcal F
 \label{blah2} \end{align}
Here $\mathbf e^{k'} \equiv S^\dag \mathbf e^{k} S$, and we suppress a left-most $S$ and right-most $S^\dag$, since $\mathcal H(\mathbf x)$ always appears (with a factor $\dx^3 \mathbf x$) under the grade projection in \eqref{foga}.

When computing expectation values of these Hamiltonians in terms of state vectors, the surviving basis transformation operators all then properly act on $(\ClR,\mathscr H)$ of which $\hat{\mathcal F} \ket\Psi$, $\check{\mathcal W}_k\hat{\mathcal F} \ket\Psi$, and $\check{\mathcal F}\hat{\mathcal W}_k \ket\Psi$ are elements:
\begin{align}
\check{\mathcal F}^\dag S^\dag \varsigma^k \overset\rightarrow\partial_k S \hat{\mathcal F} \ket\Psi& \quad (\mathcal H_+) \\
\check{\mathcal F}^\dag S^\dag \varsigma^k S^2 \overset\rightarrow\partial_k S^\dag \hat{\mathcal F} \ket\Psi& \quad (\mathcal H_-) \\
\check{\mathcal F}^\dag S^\dag \varsigma^k S \check{\mathcal F} \hat{\mathcal W}_k \ket\Psi& \vphantom{\overset\rightarrow\partial_k} \\
\check{\mathcal F}^\dag S^\dag \varsigma^k S \check{\mathcal W}_k \hat{\mathcal F} \ket\Psi& \vphantom{\overset\rightarrow\partial_k}
\end{align}
While $\mathcal H_{+}$ appears to be the minimal satisfactory correction to the problem of OBTOs---conforming more or less to the form of a covariant derivative---one cannot accept $\mathcal H_+$ while rejecting $\mathcal H_-$ based only on arguments presented thus far.  
Both are well defined and consistent with our criteria for computable geometric expressions.
If \eqref{blah} are to define physical field theories and our computability criteria carry weight, then {\it all} observables in the corresponding theories must be computable---not just the Hamiltonian.  These include first and foremost the linear and angular momenta.

\subsection{Lagrangians and Linear Momenta}
In order for the Hamiltonian to remain free of OBTOs after an inhomogeneous basis rotation and subsequent boost (and for the Lagrangian density to be Lorentz invariant), $\mathcal W_k$ must be the spatial components of a four-vector $\mathcal W_\mu$.
\begin{equation} 
\mathcal W_\mu(\mathbf x) \to \begin{cases}
S(t,\mathbf x) \mathcal W_\mu(\mathbf x) S(t,\mathbf x)^{-1} - S(t,\mathbf x) \partial_\mu S(t,\mathbf x)^{-1} & \text{(Clifford basis)} \\
\varLambda^\mu_{\phantom \mu {\mu'}} \mathcal W_{\mu}(\varLambda^{-1} \mathbf x) & \text{(coordinates)} 
\end{cases} \label{Wdef2} \end{equation}
Furthermore, in order that \eqref{Wdef} remains consistent after an inhomogeneous \emph{translation}, $\mathcal W_\mu$ cannot transform as an ordinary function of the coordinates.  Under a translation $x^\mu \to x^\mu + \epsilon^\mu(t,\mathbf x)$ with $\epsilon^\mu$ small, where $S \to S - \epsilon^\nu \partial_\nu S + \cdots$, we must have
\begin{equation} \mathcal W_\mu \to \mathcal W_\mu - \epsilon^\nu \partial_\nu \mathcal W_\mu - (\partial_\mu \epsilon^\nu) \mathcal W_\nu
+ \cdots \quad \text{(infinitesimal translation)} \label{translate} \end{equation}
Both \eqref{Wdef2} and \eqref{translate} follow from the standard form of the coordinate one-form transformation: $\mathcal W_\mu \to (\partial x^{\mu'}/\partial x^{\mu}) \mathcal W_{\mu'}$.

We construct the Lagrangian
circularly as
\begin{equation} \dx L(t,\mathbf x) = \left\langle \dx^3 \mathbf x \, \mathcal L(t,\mathbf x) \right\rangle_0, \qquad \mathcal L(t,\mathbf x) = \tilde{\mathcal P}_{0} - \mathcal H \label{lagr_def} \end{equation}
where $\tilde{\mathcal P}_0$ is the fictitious temporal component of the momentum four-vector density obtained by substituting $k \to 0$ in the expression for $\mathcal P_k$, to be determined.  Corresponding to the Hamiltonian densities in \eqref{blah}, we guess
\begin{equation}
\mathcal L_\pm(t,\mathbf x) = 
\mathcal F^\dag (\dot{\mathcal F} + [\mathcal F ,\mathcal W_0] \pm \mathcal W_0 \mathcal F) - \mathcal H_\pm \end{equation}
Under an infinitesimal translation $x^k \to x^k + \epsilon^k$, $\mathcal L_+$ transforms as
\begin{align}
\mathcal L_+ &\to \mathcal F^\dag \partial_0 (\mathcal F - \epsilon^k \partial_k \mathcal F) + \mathcal F^\dag \mathcal F (\mathcal W_0 - (\partial_0 \epsilon^k) \mathcal W_k ) + \cdots \\
\mathcal L_+ &\to \mathcal L_+ - \dot\epsilon^k \!\left( \mathcal F^\dag \partial_k \mathcal F + \mathcal F^\dag \mathcal F \mathcal W_k \right) + \cdots
\end{align}
verifying the spatial components of the momentum $P_k \equiv \partial L/\partial \dot\epsilon^k$
\begin{equation} \dx P_k \equiv \left\langle\dx^3 \mathbf x \, \mathcal P_k(\mathbf x)\right\rangle_0, \quad \mathcal P_{k,\pm}(\mathbf x) = -\mathcal F^\dag \partial_k \mathcal F - 
\mathcal F^\dag [\mathcal F, \mathcal W_k] \mp \mathcal F^\dag \mathcal W_k \mathcal F \label{momentum} \end{equation}
and hence $\tilde{\mathcal P}_{0,\pm}(\mathbf x) = \mathcal F^\dag (\dot{\mathcal F} + [\mathcal F, \mathcal W_0] \pm \mathcal W_0 \mathcal F)$.
The linear momentum $P_k$ is protected from OBTOs, as it must be by the same arguments applying to $H$.
\begin{align}
\mathcal P_{k,+}(\mathbf x) &\to -\mathcal F^\dag S^\dag \partial_k (S\mathcal F) - 
\mathcal F^\dag S^\dag S \mathcal F \mathcal W_k \\
\mathcal P_{k,-}(\mathbf x) &\to -\mathcal F^\dag S^\dag S^2 \partial_k (S^\dag\mathcal F) - 
\mathcal F^\dag S^\dag S \mathcal F \mathcal W_k + 
2 \mathcal F^\dag S^\dag S \mathcal W_k \mathcal F
\end{align}

Since $\mathcal W_0$ serves no purpose in the Hamiltonian ($\mathcal H_\pm$ contains no explicit time derivatives), we tentatively
set $\mathcal W_0 = 0$, excluding it from $\mathcal H_\pm$ and the roster of dynamical degrees of freedom with representation in the state space.
Then \eqref{lagr_def} is equivalent to the usual Legendre transform with $\mathcal H$ defined as in \eqref{blah}, and $\mathbf e^1 \mathbf e^2 \mathbf e^3 \mathcal F^\dag$ the conjugate momentum to $\mathcal F$.

\subsection{Angular Momenta}

If we identify rotations of the orientation angle invoked in Noether's theorem with global, complementary transformations of the coordinates
and bases of the local Clifford algebras,
then for the sake of computing angular momentum, $\mathcal L_+$ transforms under rotations as
\begin{equation} \mathcal L_+(\mathbf x) \to \mathcal F^\dag S^\dag \partial_0 (S \mathcal F) + \mathcal F^\dag \mathcal F \mathcal W_0 - \mathcal H_+[\mathcal F,\mathcal W_\mu](\Lambda^{-1}\mathbf x) \end{equation}
with all fields evaluated at $\Lambda^{-1} \mathbf x$.  
Taking $S$ to enact an infinitesimal rotation in the 1-2 plane, with the form $S = \mathbf 1 + \tfrac12 \theta(t) \mathbf e^1 \mathbf e^2$ where $\theta(t) \ll 1$
\begin{align}
\mathcal L_+ &\to \mathcal F^\dag (\mathbf 1 - \tfrac12 \theta \mathbf e^1 \mathbf e^2) \partial_0 ((\mathbf 1 + \tfrac12 \theta \mathbf e^1 \mathbf e^2) \mathcal F) + \cdots \\
 \mathcal L_+ &\to \mathcal L_+ + \dot\theta \, \tfrac12 \mathcal F^\dag \mathbf e^1 \mathbf e^2 \mathcal F + \cdots \end{align}
The second Lagrangian density $\mathcal L_{-}$ transforms as
\begin{align}
\mathcal L_- &\to \mathcal F^\dag S \partial_0 (S^\dag \mathcal F) + \mathcal F^\dag \mathcal F \mathcal W_0 - 2 \, \mathcal F^\dag \mathcal W_0 \mathcal F - \mathcal H_-[\mathcal F,\mathcal W_\mu](\Lambda^{-1}\mathbf x) \\ 
 \mathcal L_- &\to \mathcal F^\dag (\mathbf 1 + \tfrac12 \theta \mathbf e^1 \mathbf e^2) \partial_0 ((\mathbf 1 - \tfrac12 \theta \mathbf e^1 \mathbf e^2) \mathcal F) + \cdots \\
 \mathcal L_- &\to \mathcal L_- - \dot\theta \, \tfrac12 \mathcal F^\dag \mathbf e^1 \mathbf e^2 \mathcal F + \cdots 
 \end{align}
Differentiating with respect to $\dot\theta$ while neglecting the $\theta$ dependence owing to spatial variation of $\mathcal F$, we obtain the intrinsic angular momentum.
\begin{equation} L_\pm^3 \equiv \partial L_\pm/\partial \dot\theta, \qquad \dx L_\pm^3 = \pm \tfrac12 \left\langle \dx^3 \mathbf x \, \mathcal F^\dag \mathbf e^1 \mathbf e^2 \mathcal F \right\rangle_0 + \cdots \label{spin} \end{equation}
Since $\mathbf e^1 \mathbf e^2$ comes from $S^\dag \partial_0 S$, it is not a fixed reference element but rather transforms as $\mathbf e^1 \mathbf e^2 \to S \mathbf e^1 \mathbf e^2 S^\dag$ under subsequent rotations; the spin angular momentum is computable, as is the total angular momentum by inspection.
    \subsection{Discussion}

With the modifications in \eqref{blah}, spatially inhomogeneous rotations of the Clifford bundle applied to expressions for the Hamiltonian, linear momenta, and angular momenta are computable by the criteria proposed in \S~2.
This allows one to change between holonomic bases corresponding to Cartesian and to curvilinear coordinates on the spacelike hypersurface on which the field operators have support, by acting on state vectors from the left with a sequence of linear operators.
Can one relate $\mathcal H_\pm$ to the Standard Model of particle physics?
How might one interpret the state space?

\subsubsection{Time-dependent Basis Rotations in the Hamiltonian}
Above we have included the timelike component of the Clifford connection $\mathcal W_0$ in the Lagrangian, accounting for its non-covariant transformation under time-dependent basis rotations when calculating spin; but we have excluded it from the Hamiltonian by tentatively setting $\mathcal W_0 = 0$ and leaving out the non-covariant term.
Including $\mathcal W_0$ in $\mathcal H_+$ either introduces OBTOs when the basis transformation varies in time
\begin{equation} \mathcal H_+ \to \cdots + \mathcal F^\dag \mathcal F \mathcal W_0 - \mathcal F^\dag \mathcal F S \partial_0 S^\dag \label{HplusOBTO} \end{equation}
or requires an {\it ad hoc} rule that the non-covariant term is to be added to the Lagrangian but not to the Hamiltonian when it involves a time derivative.  

One could relax the criteria sufficient for computability to require only that the Hamiltonian be free of OBTOs accounting for \emph{spatially} varying basis transformations, since the quantum state (itself a function of time in the Schr\"odinger picture) only directly informs field configurations on a spacelike hypersurface.
Applying a time-dependent basis transformation to a field operator that only has support on a spacelike hypersurface comes across as not obviously well defined.
Alternatively, the OBTO in \eqref{HplusOBTO} can be removed by a subsequent transformation
\begin{equation} \mathcal F \to \mathcal F S \qquad \mathcal W_\mu \to S^\dag \mathcal W_\mu S - S^\dag \partial_\mu S \label{gaugetrans} \end{equation}
under which $\mathcal L_+$ is invariant.

When $\mathcal W_0$ and the non-covariant term from its basis transformation are included in $\mathcal H_-$, the OBTO cannot be so removed by a transformation that leaves the other terms invariant.
\begin{equation} \mathcal H_- \to \cdots + \mathcal F^\dag \mathcal F \mathcal W_0 - \mathcal F^\dag \mathcal F S \partial_0 S^\dag - 2 \mathcal F^\dag \mathcal W_0 \mathcal F + 2 \mathcal F^\dag S \partial_0 S^\dag \mathcal F \end{equation}
Only if the terms involving $\mathcal W_\mu$ were to not produce any changes in occupation number---e.g.\ due to conservation laws---so that one could exclude such terms outright when computing $\langle \hat H_- \rangle$, could $\mathcal H_-$ be {\it de facto} free of OBTOs to the same standard as accepted in earlier sections.

\subsubsection{Effective Spin is Real Spin}

The components of $\mathcal F$ transform as Lorentz scalar, polar 3-vector, and their Hodge duals---all left invariant after rotations of $2\pi$ radians. 
Nonetheless, when coupled as in \eqref{blah}, $\mathcal F$ ``no longer'' represents an integer-spin field.  
Multivector fields governed by $\mathcal H_+$ in \eqref{blah} \emph{effectively} transform as Weyl spinor doublets,
owing to the non-covariant term coming from \eqref{Wdef} cancelling the other ``half'' of the kinetic term's intrinsic dependence on $\dx\theta/\dx t$ after a time-dependent rotation. 
Subject to $\mathcal H_-$, the extra coupling term containing $\mathbf e^k \mathcal W_k$ flips the sign of the allowed $S^\dag\partial_\mu S$ multiplying $\mathcal F$ from the left, giving it spin opposite to fields governed by $\mathcal H_+$ for the same field configuration.  

The single-particle states must transform with unitary representations of the Lorentz group, picking up a phase of $i L_\pm^3 \theta$ under rotation, where $L_\pm^3$ is the effective angular momentum about $\mathbf e^3$.
Since the linear momenta $P_{k,\pm}$ \eqref{momentum} are defined in the same way when $\mathcal W_\mu$ vanishes and the spin contributions \eqref{spin} to $L^3_{\pm}$ differ by a sign, the doublets governed by $\mathcal H_\pm$ have opposite helicities.
Thus the single-particle states corresponding to each must transform under different representations, $(\tfrac12,0)$ and $(0,\tfrac12)$, with opposite chiralities.

\subsubsection{Electroweak Resemblance}

The $SU(2)$ gauge field $W_\mu$ of the Weinberg-Salam electroweak model \cite{Weinberg:1967tq,Salam1959} couples exclusively to left-chiral fermion fields, which are organized into doublets whose components mix under the corresponding gauge transformation.
A theoretic justification for this asymmetry (and for the appearance of an $SU(2)$ gauge group in the Standard Model, more generally) has yet to be widely accepted, though a number of attempts at such have been offered (see, e.g. \cite{Hestenes:2008qj,WITTEN1981412,A08gravi-weakunification,2010arXiv1011.5633F,hestenes1986}).

The Hamiltonian $\mathcal H_+$ in \eqref{blah} is equivalent to that of a left-chiral fermion doublet in the Standard Model under the temporal gauge $W_0 \equiv 0$, if one substitutes $\mathcal W_\mu \to i g \mathbf e^a W^a_\mu$, adds a coupling to the $U(1)$ gauge field with the usual rationale,
and adopts an anti-commuting creation/annihilation operator algebra to act on a fermionic Fock space. 
To recover a more familiar notation, one can replace $\mathcal F$ with a sum over $\psi_A$'s transforming as $\psi_A \to S \psi_A$ representing the columns of the $2 \times 2$ complex representation, while taking $\varsigma^k \otimes \mathcal W_k$ to act on the $\psi_A$'s arranged into a column doublet.
The invariance of $\mathcal L_+$ under \eqref{gaugetrans} implies $SU(2)$ gauge invariance.

An opening for an $SU(2)$ gauge interpretation of spinorial wave equations in geometric algebra has been pointed out by others---notably \cite{hestenes1986}---but in a framework in which multivector fields transform under physical rotations by left-multiplication alone.
The usual arguments about gauging a global symmetry of the Lagrangian are used to justify introducing the gauge field, which transforms under rotations only through its Lorentz coordinate index.
($\mathcal H_-$ is nonsensical in that framework.)
Rather, we have insisted that multivector fields transform by the standard dual-sided multiplication rule, and that the Clifford connection $\mathcal W_\mu$ must be introduced for the sake of computability of a first order Hamiltonian for such fields in quantum theory.

If $\mathcal L_+$ describes left-chiral fermions, then $\mathcal L_-$ describes right-chiral ones.  This right-chiral Lagrangian is not invariant under \eqref{gaugetrans}, and does not correspond to any recognizable from the Standard Model.
If $\mathcal W_\mu$ is identified with the $SU(2)$ gauge field and time-dependent OBTOs could justifiably be left out of the Hamiltonian, then a necessary condition for $\mathcal H_-$ to describe a viable field theory is that it does not couple $\mathcal F$ to massless excitations of $\mathcal W_\mu$. These must be described by a gauge invariant Lagrangian, from considerations of the missing longitudinal polarization modes \cite{Weinberg:1995mt}.  (A necessary condition for compatibility with experiment is of course that any nominal coupling terms produce no interaction between right-chiral fermions and the $W^\pm$ bosons.)

\subsubsection{In Relation to Gauge Theory Gravity}
The non-covariant term in the transformation of $\mathcal W_\mu$ is added under transformations of the bases of the Clifford bundle fibers, with respect to which the components of the field values are defined.  Under transformations of the coordinates (merely relabeling of events on the manifold), only the Greek index on $\mathcal W_\mu$ transforms.
One can imagine an object $\omega_\mu$ that behaves in the opposite way, transforming covariantly with the basis of the local Clifford algebra and non-covariantly under a coordinate transformation.
\begin{equation}
\omega_\mu(\mathbf x) \to \begin{cases}
S(t,\mathbf x) \,\omega_\mu(\mathbf x) S^{-1}(t,\mathbf x) & \text{(Clifford basis)} 
\\
\varLambda^\mu_{\phantom\mu\mu'} \omega_\mu(\mathbf x') - S_{\varLambda}^{-1}(t',\mathbf x') \partial_{\mu} S_{\varLambda}(t',\mathbf x') & \text{(coordinates)} \end{cases} \end{equation}
Here $S_\Lambda$ is a rotor that, if applied to an initially holonomic basis, would realign the vector basis elements of the geometric algebra with the local directions along which the coordinates increase after the coordinate transformation.
If one demands that $\mathcal L_\pm$ be invariant after complementary coordinate transformations and local rotations of the Clifford bundle, then one must include a coupling to $\omega_\mu$.
\begin{equation} \mathcal L_\pm = \mathcal F^\dag \mathbf e^\mu \partial_\mu \mathcal F + \mathcal F^\dag \mathbf e^\mu [\mathcal F, \mathcal W_\mu] \pm \mathcal F^\dag \mathbf e^k \mathcal W_\mu \mathcal F \pm \mathcal F^\dag \mathbf e^k \omega_\mu \mathcal F \end{equation}
Under just a coordinate rotation, we then have
\begin{align}
\mathcal L_+ &\to \mathcal F^\dag \mathbf e^\mu S_\Lambda \partial_\mu (S_\Lambda^\dag \mathcal F) + \mathcal F^\dag \mathbf e^\mu \mathcal F \mathcal W_{\mu'} + \mathcal F^\dag \mathbf e^\mu \omega_{\mu'} \mathcal F \\
\mathcal L_- &\to \mathcal F^\dag \mathbf e^\mu S_\Lambda^\dag \partial_\mu (S_\Lambda \mathcal F) + \mathcal F^\dag \mathbf e^\mu \mathcal F \mathcal W_{\mu'} - 2 \mathcal F^\dag \mathbf e^\mu \mathcal W_{\mu'} \mathcal F - \mathcal F^\dag \mathbf e^\mu \omega_{\mu'} \mathcal F \end{align}
where all fields on the RHS are evaluated at $(t',\mathbf x')$.  This indicates that the same apparent spin angular momentum results from a steady time-dependent rotation of the coordinates in one sense or a rotation of the Clifford basis in the opposite sense, as it must be.

The bivector-valued connection $\omega_\mu$ then plays the same role as the rotation gauge field $\tfrac12 \Omega(\mathbf e_\mu)$ in Gauge Theory Gravity \cite{10.1007/978-94-011-2006-7_42,doi:10.1098/rsta.1998.0178}. GTG does not separate basis and coordinate transformations, citing only a transformation law incorporating both for cases when they are holonomy-preserving:
\begin{equation} \Omega(\mathbf a) \to S \Omega(\mathbf a) S^\dag - \mathbf a \mathbf e^\mu S \partial_\mu S^\dag - \mathbf e^\mu \mathbf a S \partial_\mu S^\dag \quad \text{(basis \& coordinates)} \end{equation}
Inclusion of $\omega_\mu$ in the Hamiltonian is not required on the same grounds as $\mathcal W_\mu$, since it only affects inner transformation operators; an additional assumption like that of necessary general covariance must be made to justify it.

\section{Concluding Remarks}

We have suggested criteria for expectation values of expressions in geometric algebra (grade projections of products of multivectors) to be computable in quantum theory, considered as a generalized probability theory of linear operators.
When applied to observables in a first-order free theory of elementary multivector fields with values in $\mathcal C\ell_3(\mathbb R)$, the Hamiltonian is found not to be computable according to these criteria unless one introduces coupling terms to a bivector-valued one-form that transforms non-covariantly under basis rotations applied to fibers of the Clifford bundle.

The result is a scheme containing left- and right-chiral spin-$\tfrac12$ fields, in which all of the geometric objects are left invariant by a global rotation of $2\pi$ radians, and the two chiralities couple differently to a gauge-like connection for reasons intimately tied to their geometric character.
This scheme does not plausibly fill the explanatory gaps in the origins of maximal parity violation in the Weinberg-Salam electroweak model;
but it offers an example of a first-principles framework in which chiral asymmetry provides as much conceptual cohesion as would the null hypothesis of exact mirror symmetry.

    \bibliography{reference}{}
    \bibliographystyle{unsrt}
    
    \appendix
    
    \section{Geometric Algebra}

Consider a vector space over a field $\mathbb K$ equipped with both a symmetric inner product and the exterior wedge product 
\begin{gather*}
\mathbf a \cdot \mathbf b = \tfrac12 \left( Q(\mathbf{a+b}) - Q(\mathbf a) - Q(\mathbf b) \right) \in \mathbb K \\
 \mathbf a \wedge \mathbf b = -\mathbf b \wedge \mathbf a \end{gather*}
for some quadratic form $Q$.
Adopting the product $\mathbf a \mathbf b$ that satisfies
\[ \mathbf a \cdot \mathbf b = \tfrac12 (\mathbf a \mathbf b + \mathbf b \mathbf a) \qquad \mathbf a \wedge \mathbf b = \tfrac12 (\mathbf a \mathbf b - \mathbf b \mathbf a) \]
defines a Clifford algebra.
%
%
A Clifford algebra over $\mathbb R^n$ with $Q$ the standard $\ell_2$ norm is a {geometric algebra} $\mathcal C\ell_n(\mathbb R)$, with a set of preferred orthonormal sets $\{\mathbf e_i\}$ of basis generating elements spanning $\mathbb R^n$ that satisfy
\[ \mathbf e_i \mathbf e_j + \mathbf e_j \mathbf e_i = 2\delta_{ij} \]
A generic element of $C\ell_2(\mathbb R)$ written in terms of such a basis takes the form 
\[ \mathbf a = a_0 + a_1 \, \mathbf e_1 + a_2 \, \mathbf e_2 + a_{12} \, \mathbf e_1 \mathbf e_2 \]
with real components $a_0$, $a_i$, $a_{12}$. The Pauli matrices that feature prominently in quantum mechanics provide a representation of $C\ell_3(\mathbb R)$: 
\[ \sigma_1 = \begin{pmatrix} \matzero & 1 \\ 1 & \matzero \end{pmatrix} \quad \sigma_2 = \begin{pmatrix} \matzero & -i \\ i & \matzero \end{pmatrix} \quad \sigma_3 = \begin{pmatrix} 1 & \matzero \\ \matzero & -1 \end{pmatrix} \]
There are countless other equally valid matrix representations, e.g.\
\[ \begin{pmatrix} \matzero & 1 & 0 & 0 \\ 1 & \matzero & 0 & 0 \\ 0 & 0 & \matzero & -i \\ 0 & 0 & i & \matzero \end{pmatrix}, \quad
 \begin{pmatrix} \matzero & 0 & 1 & 0 \\ 0 & \matzero & 0 & i \\ 1 & 0 & \matzero & 0 \\ 0 & -i & 0 & \matzero \end{pmatrix}, \quad
\begin{pmatrix} \matzero & 0 & 0 & 1 \\ 0 & \matzero & -i & 0 \\ 0 & i & \matzero & 0 \\ 1 & 0 & 0 & \matzero \end{pmatrix} \]

If a multivector can be expressed as a linear combination of $(k-1)$-fold exterior products of vector basis elements, then it has a well defined {\it grade} of $k$ (0 if the multivector is just a real number) and is called a {\it $k$-blade}.  In that case the {\it grade-$k$ projection} is equal to the original multivector, and all other grade projections are equal to 0.
The matrix representation of the unit grade-0 element is always the identity matrix.
The grade-0 projection may be computed as the trace of the matrix representation divided by its dimension.
\[ \langle \mathbf a \rangle_0 = \tfrac1d \Tr A \]
Likewise, any higher grade projection can be computed in terms of the trace of a product with each of a complete basis of orthonormal elements with that grade, e.g.\
\[ \textstyle \langle \mathbf a \rangle_1 = \sum_i \mathbf e^i \left\langle \mathbf a (\mathbf e^i)^\dag \right\rangle_0 = \tfrac1d \sum_i \varsigma^i \Tr \left[ A (\varsigma^i)^\dag \right] \]
where ${}^\dag$ denotes the {\it reverse} involution of a multivector
\[ (\mathbf a \mathbf b)^\dag \equiv \mathbf b^\dag \mathbf a^\dag \qquad \mathbf 1^\dag \equiv \mathbf 1 \qquad (\mathbf e^i)^\dag \equiv \mathbf e^i \]

Multivectors transform under rotations by the action of {\it rotors}
\[ S \in \left\{ \exp(\tfrac14 \epsilon_{ijk} \theta_i \mathbf e^j \mathbf e^k) \mid \theta_i \in \mathbb R \right\} \]
as $\mathcal F \to S \mathcal F S^{-1}$ (or perhaps $\psi \to S \psi$).  Since $S^\dag S= S S^\dag = \mathbf 1$ for rotations, the matrix representation of $S$ is unitary, and we often write the transformation as $\mathcal F \to S \mathcal F S^\dag$ in the case of rotations.  
If the grade projections of $\mathcal F$ are taken to be Lorentz scalar, polar 3-vector, and their Hodge duals, then a Lorentz boost is performed by substituting for the bivector $S$ the paravector
\[ \mathcal S \in \left\{ \exp(\tfrac12 w_i \mathbf e^k) \mid w_i \in \mathbb R \right\} \]
This boost paravector is its own reverse: $\mathcal S = \mathcal S^\dag$; its inverse is obtained by taking $w_i \to -w_i$.  The 3-volume element $\dx^3 \mathbf x$ is the time-like component of a 4-pseudovector, transforming under boosts as
\[ \dx^3 \mathbf x \to \mathcal S \, \dx^3 \mathbf x \, \mathcal S^\dag = \mathcal S \, \dx^3 \mathbf x \, \mathcal S \]
    
\end{document}